\documentclass[aps,pra,reprint,physrev,twocolumn,groupedaddress]{revtex4-2}

\usepackage{graphicx}
\usepackage{dcolumn}
\usepackage{bm}
\usepackage{amsmath}
\usepackage{amssymb}
\usepackage{braket}
\usepackage{siunitx}
\usepackage[T1]{fontenc}
\usepackage{xcolor}

\usepackage{tikz}
\usetikzlibrary{decorations.pathmorphing,arrows.meta,calc,positioning}

\usepackage[colorlinks]{hyperref}
\usepackage[noabbrev]{cleveref}
\crefname{equation}{Eq.}{Eqs.}
\crefrangelabelformat{equation}{(#3#1#4--#5#2#6)}

\graphicspath{{.},{./fig/}}
\DeclareGraphicsExtensions{.pdf,.png,.eps}

\begin{document}

\preprint{APS/123-QED}

\title{Quantum open system description of a hybrid plasmonic cavity}

\author{Marco Vallone}
\email{marco.vallone@polito.it}
\affiliation{Dipartimento di Elettronica e Telecomunicazioni, Politecnico di Torino,\\
Corso Duca degli Abruzzi 24, 10129 Torino, Italy}

\date{\today}

\begin{abstract}
We present a unified quantum open system framework for lossy plasmonic cavities in which coherent dynamics, relaxation, dephasing, and irreversible absorption are treated on equal footing. The Dyson equation for the cavity photon propagator in the random-phase approximation yields a complex self-energy $S(\omega)$ that accounts for both the renormalization and the damping of hybrid plasmon–photon modes (polaritons, in a quasi-particle description). Tracing out the electronic and photonic environments leads to a Liouvillian for the upper (UP) and lower (LP) polaritonic branches, incorporating leakage through $\Gamma=-2\,\mathrm{Im}\,S(\omega)$, internal UP$\leftrightarrow$LP scattering rates, and dephasing. Time evolution equations for polariton populations, interbranch coherence, and driven amplitudes in closed form also provide analytic expressions for their steady-state values, the quench rate of UP–LP oscillations and polaritonic lineshapes, valid in the limit of low polaritonic density, but covering light-matter ultrastrong coupling. The theory establishes a self-consistent description of dissipative polariton dynamics in plasmonic and nanophotonic cavities, directly applicable to response spectra, time-domain measurements, and dissipation engineering.
\end{abstract}

\maketitle

\section{Introduction}
\label{s:introduction}

\begin{figure}[!t]
\centering
\includegraphics[width=0.9\columnwidth]{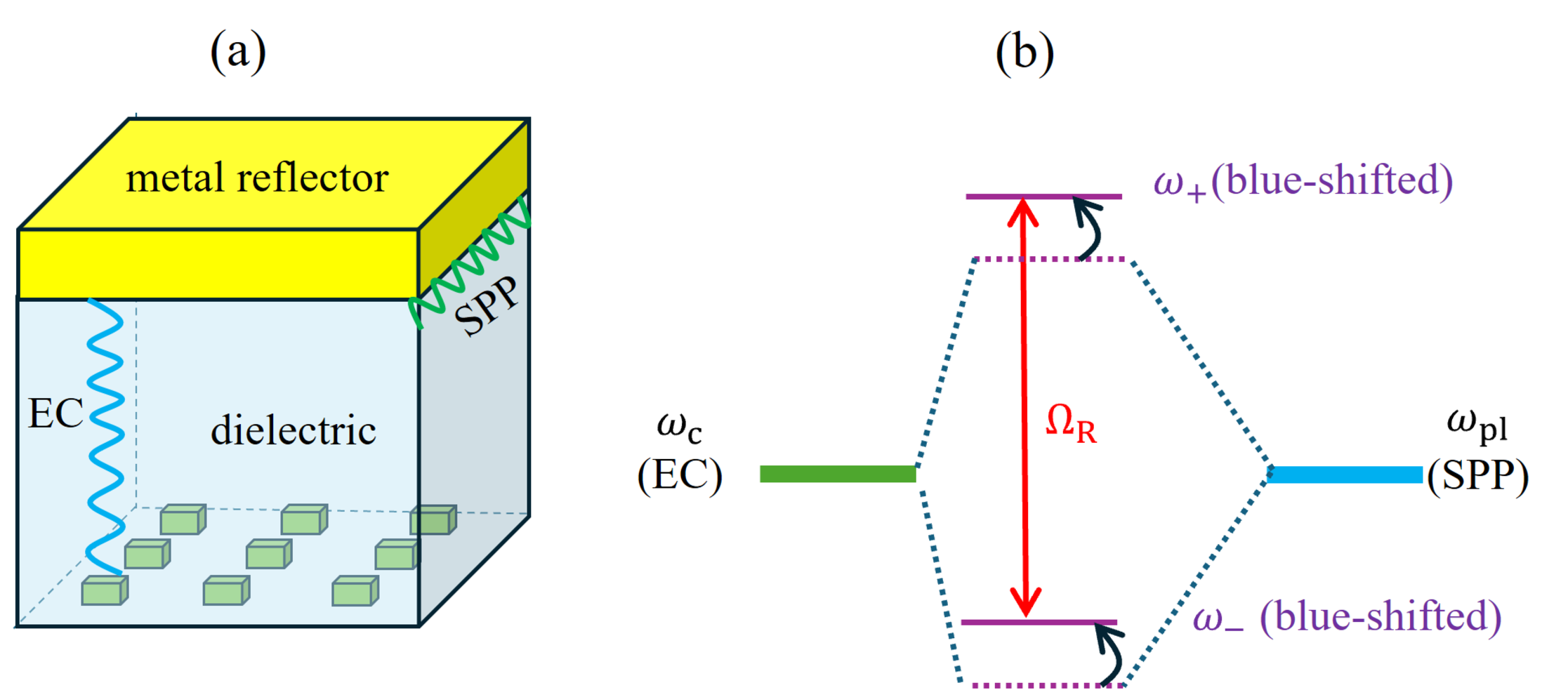}
\includegraphics[width=0.9\columnwidth]{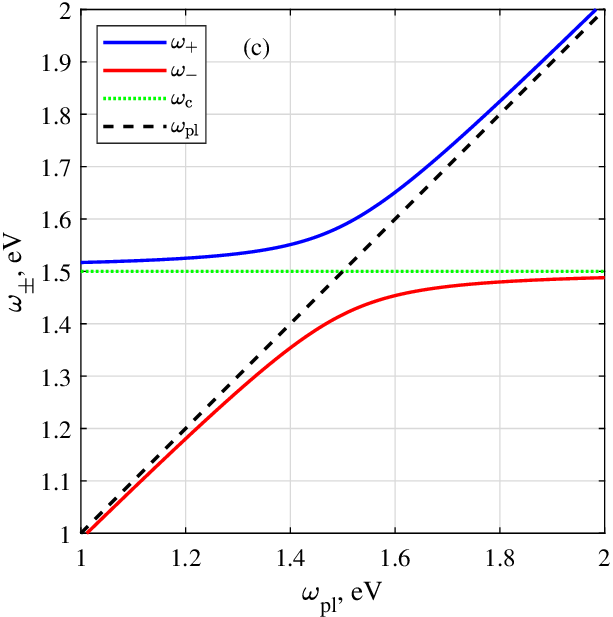}
\caption{(a) Sketch of a resonant plasmonic cavity illuminated at normal incidence. A lattice of gold nanoparticles on the illuminated face excites an SPP mode with frequency $\omega_\mathrm{pl}$, set by the lattice period, that propagates along the dielectric/reflector (assumed gold) interface, while the cavity supports an EC mode at $\omega_\mathrm{c}$. (b) Schematic representation of EC--SPP hybridization and (c) dispersion of eigenfrequencies $\omega_\pm$ for $\omega_\mathrm{c} = 1.5$\,eV, plasma frequency $\Omega_\mathrm{pl} \approx 8.5$\,eV, adimensional EC-SPP coupling constant $\eta = 0.02$, and zero losses, \cref{eq:H_QED_total,eq:polaritonic_Hamiltonian,eq:QED_eigenvalues}.}
\label{f:fig_1}
\end{figure}

Confinement and modulation of light in resonant cavities enable access to the regime of extreme light-matter interactions for fundamental studies and the development of highly compact nanophotonic devices. Plasmonic nanoparticles, periodically assembled as shown in Fig.~\ref{f:fig_1}(a), can enhance local electromagnetic fields and trigger collective electronic excitations, such as surface plasmon polaritons (SPP), in metallic or highly doped layers of the cavity itself \cite{1988Raether,Ozbay2006,Maier2007}. Strong coupling between SPP and electromagnetic cavity (EC) modes eventually results in their hybridization, giving rise to upper (UP) and lower (LP) polaritons with Hopfield-like dispersion \cite{1958Hopfield_PR} as shown in Fig.~\ref{f:fig_1}(b,c). Their characteristics are exploited in photonic devices for sensing, photodetection, and nanophotonic applications \cite{Atwater2010,Berini2014,Anker2008,Homola1999,Catchpole2008,Polman2012,2023Sarkar_ASR}.

Cavity quantum electrodynamics (cQED) provides a rigorous framework for light--matter interactions in the quantum regime~\cite{1992CohenTannoudji}. In the weak-coupling limit, the Jaynes--Cummings model describes coherent excitation exchange under the rotating-wave approximation (RWA) \cite{1963JaynesCummings_PIEEE,2013Hummer_PRB,2014LiuJ_OEX,2024Larson}, while the quantum Rabi
model \cite{1936Rabi_PR,1937Rabi_PR} becomes essential in the ultrastrong and
deep-strong coupling regimes \cite{2005Ciuti_PRB,2019FornDiaz_RMP,2019Kockum_NATRP,2020Baranov_NATC}. Strong coupling also renormalizes the bare cavity response. Experimental dispersions often exhibit a blue shift of the polaritonic branches, which has been shown to arise from the dressing of the photon propagator in the cavity by a real-valued self-energy $\Sigma(\omega)$, which represents SPP-SPP interactions \cite{2024Vallone_OEX,2025Vallone_NPJ}. 

A central challenge is dissipation, arising from Ohmic losses, radiative leakage, and interfacial scattering, which reduce coherence and limit accessible light-matter coupling, especially in metal. However, several material and geometric strategies now enable substantially lower plasmonic damping: among others, atomically smooth single-crystalline Au \cite{2010Huang_NATC,2022Liu_NL} and Ag films \cite{2020Baburin_Coatings}, graphene-plasmon gratings \cite{2023Guo_JAP}, Na-based nanostructures \cite{2023Rawashdeh_NL}, plasmonic materials based on C-boron nitride monolayers \cite{2024Gorashi_PRM}, graphene multilayers and metamaterial waveguides \cite{2014Qin_OEX,2025Asadi_OPTIK}. These examples demonstrate that underdamped plasmonic cavities are experimentally feasible, motivating a self-consistent open system treatment of hybrid plasmon-photon dynamics.

When losses are considered, the self-energy becomes complex,  
\begin{equation}
    \mathcal{S}(\omega)=\Sigma(\omega)-\mathrm{i}\,\Gamma(\omega)/2 ,
\label{eq:Sigma_definition}    
\end{equation}
where $\Gamma(\omega)=-2\,\mathrm{Im}\,S(\omega)$ describes irreversible coupling to electronic continua~\cite{2007Breuer}.
In this work, we combine Hopfield diagonalization with the renormalization of the photon propagator in an electrons plasma to obtain the eigenfrequencies of a hybrid plasmonic cavity renormalized by a complex self-energy (Sec.\,\ref{s:Dyson}). Then, we derive a Gorini--Kossakowski--Sudarshan--Lindblad (GKSL) master equation \cite{1976Lindblad_CMP,1976Gorini_JMP,2007Breuer} for UP and LP polaritons that includes leakage, UP$\leftrightarrow$LP thermalization, and dephasing (Sec.\,\ref{s:open_system}). Closed evolution equations for coherent amplitudes, populations, and interbranch coherence yield analytic steady-state solutions and explain the quenching of UP–LP oscillations under coherent Raman-like driving (Sec.\,\ref{s:dissipative_dynamics}). Section \ref{s:conclusions} summarizes the results and discusses possible extensions to non-Markovian environments.

Throughout this work we use natural units ($\hbar=c=k_\mathrm{B}=1$, which indicate the reduced Planck constant, the light velocity in vacuum, and the Boltzmann constant, respectively) and report frequencies and energies in eV.

\section{Lossy plasmonic cavity: Hamiltonian and propagator formalisms}
\label{s:Dyson}
We refer to the resonant plasmonic cavity sketched in Fig.~\ref{f:fig_1}(a) and illuminated at normal incidence by an electromagnetic field. A lattice of gold nanoparticles on the illuminated face excites an SPP mode with frequency $\omega_\mathrm{pl}$, set by the lattice period, that propagates along the dielectric/reflector interface, while the cavity supports an EC mode at $\omega_\mathrm{c}$. The reflector is assumed to be gold.

There are several excellent papers in the literature that describe the EC-SPP interaction in the context of the cQED \cite{1958Hopfield_PR,2012Todorov_PRB,2020Mueller_NAT,2023WangZ_JAP}, in which the interaction produces hybrid modes with angular frequencies $\omega_\pm$.
When modeled as an isolated system (i.e., neglecting any losses), the Hamiltonian in the Power-Zienau-Woolley (PZW) multipolar picture \cite{1957Power_NC,1971Woolley_PRSLA,2012Todorov_PRB} is
\begin{align}
\hat{H}_{\mathrm s} &= 
\omega_{\mathrm c}\!\left(\hat a^{\dagger}\hat a+\tfrac{1}{2}\right)
+\omega_{\mathrm{pl}}\!\left(\hat b^{\dagger}\hat b+\tfrac{1}{2}\right) \nonumber \\
&-\frac{\mathrm{i}\eta\,\Omega_{\mathrm{pl}}}{2}
(\hat a-\hat a^{\dagger})(\hat b+\hat b^{\dagger})
+\frac{\eta^{2}\Omega_{\mathrm{pl}}^{2}}{2}
(\hat b+\hat b^{\dagger})^{2},
\label{eq:H_QED_total}
\end{align}
where $\hat{a}$ and $\hat{b}$ are bosonic annihilation operators for EC and SPP modes, respectively, $\Omega_{\mathrm{pl}}$ is the bulk plasma frequency, and $\eta\in[0,1]$ is a dimensionless EC--SPP coupling factor that depends on the cavity details. It has been demonstrated \cite{2012Todorov_PRB,2021Vukics_SREP} that this formulation of $\hat{H}_{\mathrm s}$ is gauge invariant and equivalent to the Hopfield Hamiltonian \cite{1958Hopfield_PR}. Both include counter-rotating terms, and the quadratic term $(\hat b+\hat b^{\dagger})^{2}$ is equivalent to the usual $\hat{A}_\mu \hat{A}^\mu$ term, which arises from minimal coupling (where $\hat{A}^\mu$ is the electromagnetic vector potential operator) and is responsible for a blue shift of the Hamiltonian eigenfrequencies, as shown in Fig.\,\ref{f:fig_1}(b). This blue shift is experimentally observed when the EC-SPP coupling is sufficiently strong \cite{2020Baranov_NATC,2021Yoo_NATP,2023Pisani_NATC}. Following Ref.~\cite{2025Vallone_NPJ}, $\hat{H}_{\mathrm s}$ can be diagonalized by a Hopfield-Bogoliubov transform to obtain
\begin{equation}
    \hat{H}_{\mathrm s}= \sum_{k=\pm} \omega_k \hat{B}_k^\dagger \hat{B}_k = \omega_{+} \hat{B}_{+}^\dagger \hat{B}_{+} + \omega_{-} \hat{B}_{-}^\dagger \hat{B}_{-},
\label{eq:polaritonic_Hamiltonian}
\end{equation}
where $\hat{B}_\pm$ are bosonic annihilation operators for Fock states $\ket{n_\pm}$, and the eigenvalues are
\begin{equation}
\omega_{\pm}^2 = \frac{\omega_\mathrm{c}^2 + \widetilde{\omega}^2_\mathrm{pl}}{2} \pm  \frac{\sqrt{\left(\omega_\mathrm{c}^2 - \widetilde{\omega}^2_\mathrm{pl}\right)^2 + 4 g^2 \omega_\mathrm{c}^2}}{2},
\label{eq:QED_eigenvalues}
\end{equation}
with coupling strength $g \equiv \eta\,\Omega_\mathrm{pl}$ and $\widetilde{\omega}_\mathrm{pl} \equiv \sqrt{\omega^2_\mathrm{pl} + g^2}$. 
The dispersion as a function of $\omega_\mathrm{pl}$ is shown in Fig.~\ref{f:fig_1}(c) for $\omega_\mathrm{c} = 1.5$\,eV, $\Omega_\mathrm{pl} \approx 8.5$\,eV (typical for gold \cite{2012Olmon_PRB}), and $\eta = 0.02$, which yields $g \approx 0.17$\,eV.

The energy separation between the modes $\omega_{+}-\omega_{-}$ has a minimum at the crossing, which can be found taking its derivative with respect to $\omega_\mathrm{pl}$, obtaining 
\begin{equation}
    \min\left(\omega_{+}-\omega_{-}\right) = g ,
\end{equation}
which defines the Rabi frequency as $\Omega_\mathrm{R}= g$. The coupled EC-SPP system behaves as two coupled harmonic oscillators, and $\ket{n_\pm}$ are the corresponding bosonic normal modes. The description given so far is valid in the absence of losses, from the weak to the ultrastrong coupling regimes, the latter defined as in Ref.~\cite{2005Ciuti_PRB}, \textit{i.e.}, when $\Omega_\mathrm{R}$ approaches $\min\{\omega_\mathrm{c},\omega_\mathrm{pl}\}$. 

\subsection{Introduction of losses: the retarded propagator formalism}
\label{s:Green}
The spatial components of the retarded free photon propagator (Green's function)
\begin{equation}
D_{0,\mu\nu}^+(x-y)=\bra{ 0}\mathrm{T}\,\hat{A}_\mu(x) \hat{A}_\nu^\dag(y) \ket{0}
\end{equation}
(where $\mathrm{T}$ denotes the time-ordering symbol), in conjugate space is given by  
$D_{0,ij}(\omega, \mathbf{k}) = \delta_{ij}\, \omega^2 / (\omega_\mathbf{k}^2 - \omega^2 + \mathrm{i}0^+)$ \cite{1995Peskin}. The photon propagator $D_{ij}$, which describes the propagation of a photon in the cavity, satisfies the Dyson equation,
\begin{equation}
D_{ij} = D_{0,ij} + D_{0,ik}\, \Pi^{kl}\, D_{lj},
\end{equation}
where the polarization operator $\Pi^{kl}$ describes the interactions of EC photons with the medium.

\begin{figure}[t]
\centerline{\includegraphics[width=1\columnwidth]{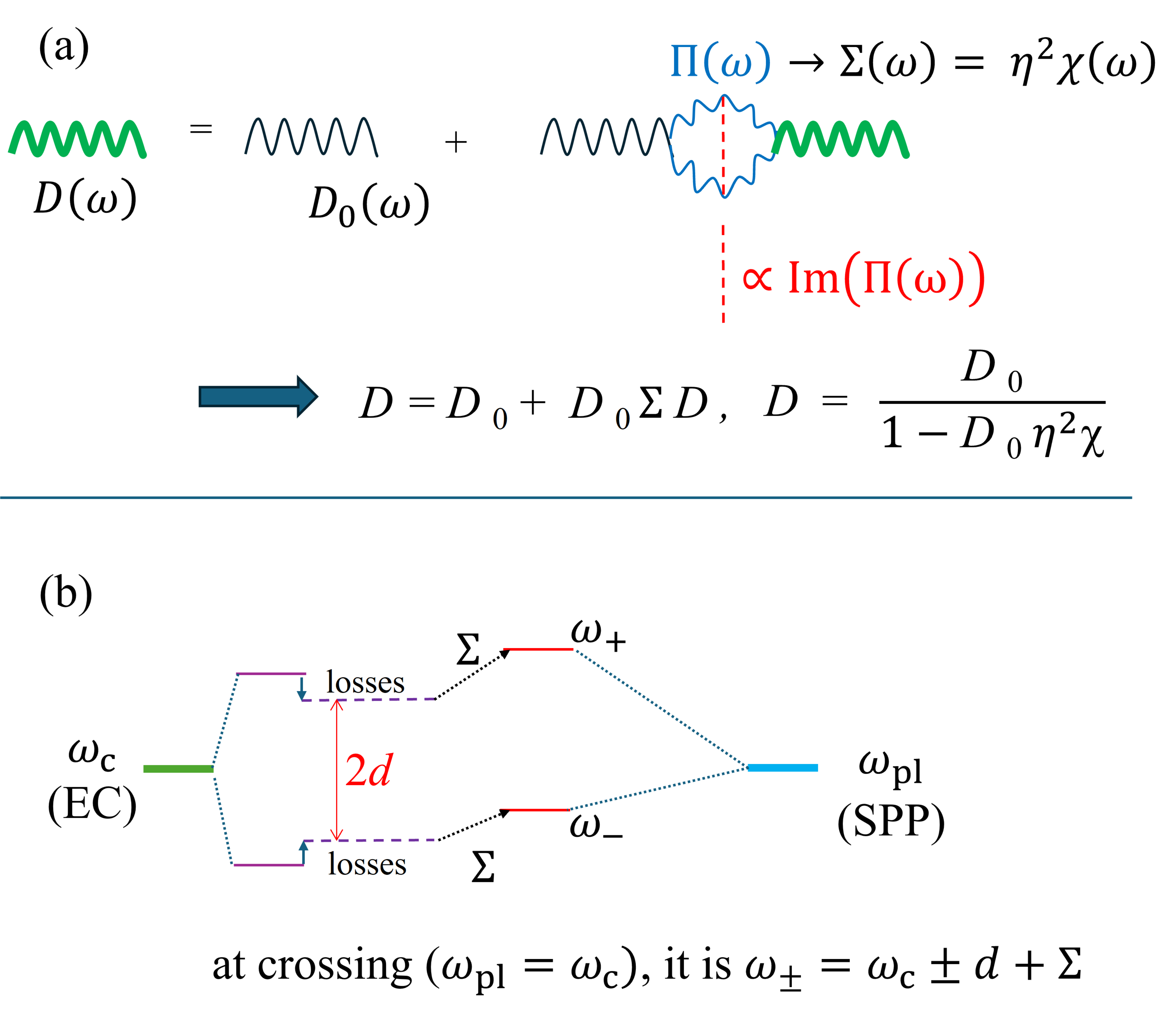}}
\caption{(a) Diagrammatic view of the Dyson equation for $D$, with amplitude damping represented as the imaginary part of the self-energy. The free photon propagator $D_0$ is dressed by the polarization bubble $\Pi(\omega)$. The vertical dashed \emph{cut} through the bubble represents on-pole intermediate states and yields $2\,\mathrm{Im}\,\mathcal{S}(\omega)$ (optical theorem \cite[Ch.~4]{1995Peskin}), \textit{i.e.}, physical loss channels (absorption in the metal). (b) EC–SPP hybridization in the strong-coupling regime with losses, which reduce (and possibly set to zero) the Rabi frequency splitting.}
\label{f:fig_2}
\end{figure}

When dominant EC and SPP modes are present, $D_{0,ij}(\omega,\mathbf{k})$ becomes the scalar function $D_0(\omega) = \omega^2 / (\omega_\mathrm{c}^2 - \omega^2 + \mathrm{i}0^+)$.  
The Dyson equation then reduces to $D = D_{0} + D_{0}\,\Pi\,D$, with $\Pi(\omega)$ proportional to the electric susceptibility of the medium, $\chi(\omega)$, \textit{i.e.}, $\Pi(\omega) \equiv \mathcal{S}(\omega) = \eta^2 \chi(\omega)$, valid in the long-wavelength limit of the random-phase approximation (RPA) \cite{2020Rukelj_NJP,2024Hughes_OPTQ}.  
For $\chi(\omega)$ we adopt the Drude--Lorentz form \cite{1976Ashcroft,1988Raether,Maier2007},
\begin{equation}
\chi(\omega) = \frac{\Omega_{\mathrm{pl}}^2}
{\omega_{\mathrm{pl}}^2 - \omega^2 - \mathrm{i}\gamma_{\mathrm{D}}\omega},
\label{eq:chi_Drude}
\end{equation}
where $\gamma_{\mathrm{D}}$ accounts for SPP absorption, and Fig.~\ref{f:fig_2}(a) schematically illustrates the mechanism.

Without loosing generality, all irreversible losses (or leakage) is meant to come from the modes absorption in the metallic reflector, described by the imaginary part of $\chi(\omega)$, assuming for the gold reflector $\gamma_\mathrm{D} \approx 0.05$\,eV, \cite{2012Olmon_PRB}. Additional, independent leakage channel (\textit{e.g.}, finite mirror reflectivity, out-coupling to free space, roughness) although not explicitly included in the present description, can be considered as well without any difficulty (for example, the radiative losses at cavity boundaries -- the mirrors -- can be described by introducing a factor $-\mathrm{i} \kappa \omega$ instead of the $+0^+$ in the expression of the free photon propagator $D_0(\omega)$, where $\kappa$ is the bare photon loss rate at mirrors).

The solution to the Dyson equation is
\begin{align}
D(\omega) &= \frac{1}{D_0^{-1}(\omega) - \mathcal{S}(\omega)}  \nonumber \\ 
&= \frac{\omega^2 (\omega^2_\mathrm{pl} - \omega^2 - \mathrm{i}\gamma_\mathrm{D}\omega)} 
{(\omega_{\mathrm{c}}^2 - \omega^2)(\omega^2_\mathrm{pl} - \omega^2 - \mathrm{i}\gamma_\mathrm{D}\omega)
- \omega^2g^2}.
\label{eq:D_omega}
\end{align}
Neglecting leakage ($\gamma_\mathrm{D}=0$), the four poles of $D(\omega)$ are the symmetric $\pm\omega_\pm$, with $\omega_\pm$ given by \cref{eq:QED_eigenvalues}, confirming the equivalence of Hamiltonian and propagator formalisms \footnote{The four poles represent two physical modes, the LP and the UP; the positive frequencies $\omega_\pm$ are the corresponding mode energies, while the two poles at $-\omega_\pm$ are required by reality and causality.}. 
\begin{figure*}[!t]
\centerline{
\includegraphics[width=0.33\textwidth]{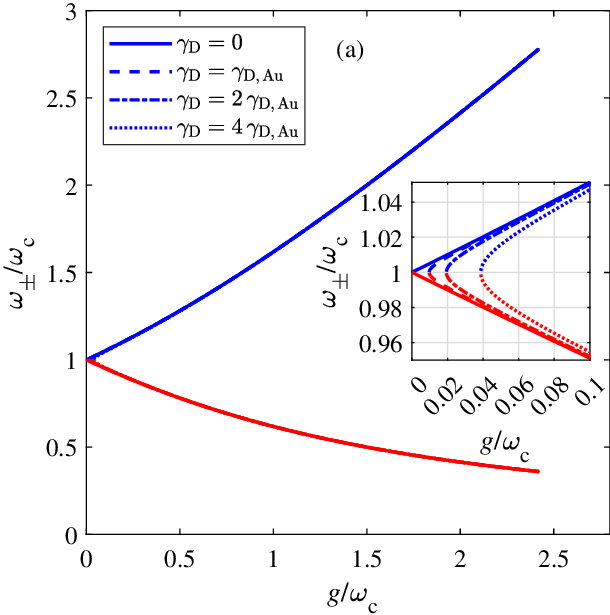}
\includegraphics[width=0.33\textwidth]{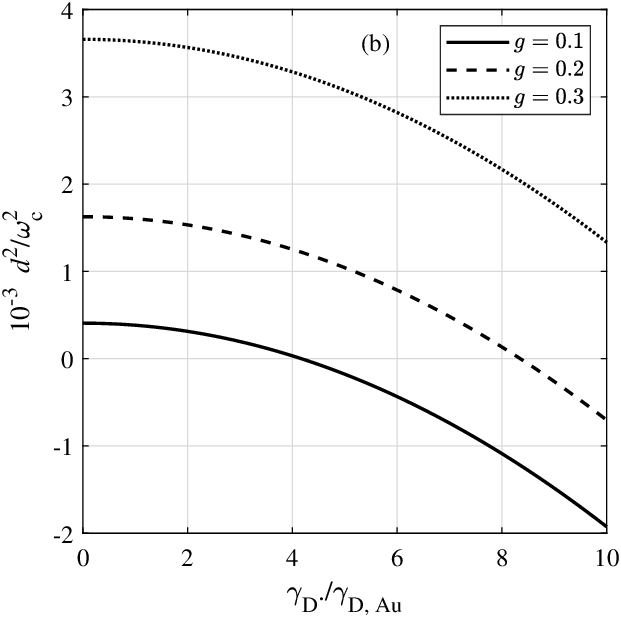}
\includegraphics[width=0.33\textwidth]{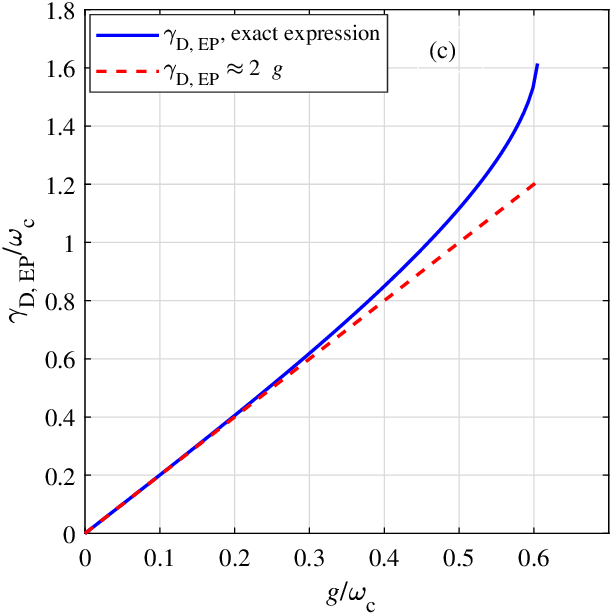}}
\caption{(a) The eigenvalues $\omega_\pm$ as function of the normalized coupling strength $g/\omega_\mathrm{c}$. The inset shows that, in presence of losses, the polaritonic splitting does not exist for very low values of $g/\omega_\mathrm{c}$. (b) The normalized square polaritonic energy separation as function of the loss rate $\gamma_\mathrm{D}$ normalized to the value for gold, for three values of $g$. (c) The loss rate $\gamma_\mathrm{D,\,EP}$ that defines the exceptional point, above which the polaritonic splitting cannot exist, plotted against normalized coupling strength.}
\label{f:fig_3}
\end{figure*}
%
\subsection{Effect of irreversible leakage at crossing}
\label{s:losses_irreversible}
At the resonant condition $\omega_\mathrm{pl}=\omega_\mathrm{c}$ (crossing), one can work from either the Hamiltonian or propagator side. However, the latter is convenient here because the impact of leakage appears directly in the denominator of \cref{eq:D_omega}.

To obtain the eigenfrequencies at crossing, we equate the denominator of \cref{eq:D_omega} to the product $(\omega-\omega_{+})(\omega+\omega^*_{+})(\omega-\omega_{-})(\omega+\omega^*_{-})$ and require
\begin{equation}
\omega_\pm = \omega_\mathrm{c} \pm d + \Sigma - \mathrm{i}\Gamma/2,
\label{eq:eigenfreq}
\end{equation}
where $d$ quantifies the mode splitting, $\Sigma=\Re[\mathcal{S}]$ is the energy renormalization, and $\Gamma=-2\Im[\mathcal{S}]$ defines the linewidth. Matching the coefficients of the characteristic polynomial yields the analytic expressions (valid at crossing),
\begin{align}
d^2 &= \frac{\omega_\mathrm{c}^2}{2} + \frac{g^2}{4} -\frac{3}{32}\gamma_\mathrm{D}^2  \nonumber \\
&\quad - \frac{1}{8}\sqrt{16\,\omega_\mathrm{c}^4 - \gamma_\mathrm{D}^2\!\left(2\omega_\mathrm{c}^2+g^2+\frac{5}{16}\gamma_\mathrm{D}^2\right)} ,
\label{eq:d2}
\end{align}
\begin{align}
\Sigma &= -\omega_\mathrm{c} 
+ \frac{1}{2}\!\left[2\omega_\mathrm{c}^2+g^2-\frac{3}{8}\gamma_\mathrm{D}^2 \right. \nonumber \\
&\quad \left. +2\!\sqrt{\omega_\mathrm{c}^4 
- \frac{1}{16}\!\left[\left(2\omega_\mathrm{c}^2+g^2\right)\gamma_\mathrm{D}^2 
+ 5\gamma_\mathrm{D}^4\right]}\right]^{1/2} , \nonumber \\[2mm]
\Gamma &= \gamma_\mathrm{D}/2 .
\label{eq:SigmaGamma}
\end{align}
The Rabi frequency,
\begin{equation}
\Omega_\mathrm{R} = (\omega_{+} - \omega_{-})_{\omega_\mathrm{pl}=\omega_\mathrm{c}} = 2 d,
\end{equation}
is set entirely by $d$ (the common shift $\Sigma$ cancels). In the lossless limit, we recover $d\to g/2$ and $\Omega_\mathrm{R}\to g$.

For sufficiently large losses, $d^2$ becomes negative, signalling an exceptional point (EP) and the loss of a real normal-mode splitting, as shown in Fig.\,\ref{f:fig_3}(a,b). The value of the loss rate $\gamma_\mathrm{D,\,EP}$ above which this happens is determined by the condition $d^2=0$, which yields
\begin{equation}
    \gamma_\mathrm{D,\,EP}^2 =\frac{1}{7}\left(8 g^2 + 16\omega^2_\mathrm{c} -4 \sqrt{16 \omega^4_\mathrm{c} - 40 g^2\omega^2_\mathrm{c} -10g^4} \right) . 
\label{eq:EP}
\end{equation}
Fig.\,\ref{f:fig_3}(c) shows $\gamma_\mathrm{D,\,EP}$  as function of $g/\omega_\mathrm{c}$ according to \cref{eq:EP}, with the compact estimate $\gamma_\mathrm{D,EP} \approx 2\,g$ valid in the experimentally relevant regime $\{g,\gamma_\mathrm{D}\}\ll\omega_\mathrm{c}$. A well-working plasmonic cavity has $\gamma_\mathrm{D} \ll \gamma_\mathrm{D,EP}$, since for $\gamma_\mathrm{D}>\gamma_\mathrm{D,EP}$ the hybrid modes are distinguished by their decay rates rather than by their resonance frequencies.

Even in case the influence of losses on the dispersion is small, yet their inclusion is essential for a consistent open system description of the plasmonic cavity.  
In Section\,\ref{s:open_system}, the full complex self-energy $\mathcal{S}(\omega)$ is shown to enter the open system dynamics through the density operator: although its real part $\Sigma(\omega)$ only renormalizes the Hamiltonian and can be removed from the dynamics, its imaginary part $-\Gamma/2$ defines the irreversible damping channel that contributes a non-Hermitian term to the Liouvillian and directly affects the dynamics. This establishes a self-consistent link between microscopic electron response, the Green's function description of the cavity, and density-matrix dynamics.  
Also UP$\leftrightarrow$LP relaxation/excitation processes (\textit{e.g.}, phonon-assisted \cite{2011Virgili_PRB}) are shown to determine the dynamics of the system and will be described in a similar way by means of a separate contribution to the Liouvillian of the system.

\section{The cavity as open system}
\label{s:open_system}

When losses are zero, \cref{eq:eigenfreq,eq:d2,eq:SigmaGamma} yield the same results as \cref{eq:QED_eigenvalues} at crossing. However, in the presence of irreversible cavity losses characterized by $\gamma_\mathrm{D}$ and restricting the description to the crossing, the propagator formulation described in Sec.~\ref{s:Dyson} most transparently captures the system's physics. Irreversible absorption, UP$\leftrightarrow$LP relaxation/excitation processes \cite{2011Virgili_PRB}, and dephasing are addresses together in this section as dissipation processes relevant for cavity dynamics, within the density operator formalism for quantum open systems \cite{2007Breuer,2020Manzano_ADV}.

To this end, the hybrid cavity can be conveniently described as a bosonic two-mode system in the Fock polaritonic basis $\ket{n_{+}, n_{-}},\,\,n_\pm = 0, 1, 2,\ldots$ (the vacuum corresponds to $n_{+}= n_{-}= 0$). The Fock states $\ket{\mathbf{n}} \equiv \ket{n_{+}, n_{-}}$, generated from the vacuum by $B_\pm^\dag$ are defined as
\begin{align}
    \ket{\mathbf{n}} &=\frac{(\hat{B}_{+}^\dagger)^{n_{+}}(\hat{B}_{-}^\dagger)^{n_{-}}}{\sqrt{n_{+}!\,n_{-}!}}\;\ket{0_{+},0_{-}} \nonumber \\
\hat{B}_\pm\ket{\mathbf{n}} &= \sqrt{n_\pm}\,\ket{\mathbf{n}-\mathbf{e}_\pm},
\label{eq:Fock_states}
\end{align}
with $\mathbf{e}_{+}=(1,0)$ and $\mathbf{e}_{-}=(0,1)$. They belong to the Hilbert space of the isolated system -- the system of interest -- for which $\{\ket{n_{+},n_{-}}\}$ is an orthonormal basis in which $\hat{H}_\mathrm{s}$ is diagonal, $\hat{H}_\mathrm{s}=\omega_{+}\,\hat{B}_{+}^\dagger\hat{B}_{+} + \omega_{-}\,\hat{B}_{-}^\dagger\hat{B}_{-}$, and has eigenvalues $E_\mathbf{n} \equiv E_{n_{+},n_{-}}=n_{+}\omega_{+} + n_{-}\omega_{-}$.

A vector state $\ket{\phi_k}$ of the Hilbert space associated with the total system, defined as the isolated system and the environment, evolves according to the von Neumann equation $\dot{\hat{\rho}}_\mathrm{T} = -\mathrm{i}[\hat{H}_\mathrm{tot}, \hat{\rho}_\mathrm{T}]$, where $\hat{\rho}_\mathrm{T}=\sum_k\ket{\phi_k}\bra{\phi_k}$ is the total density operator and $\hat{H}_\mathrm{tot} = \hat{H}_\mathrm{s} + \hat{H}_\mathrm{env} + \hat{H}_\mathrm{I}$ is the total Hamiltonian. Here, $\hat{H}_\mathrm{env}$ is the environment -- or bath -- Hamiltonian and $\hat{H}_\mathrm{I}$ describes the system-bath interactions). Tracing over the environment degrees of freedom, the reduced density operator $\hat{\rho}$ is defined as
\begin{align}
\hat{\rho}&=\sum_{\mathbf{n},\mathbf{m}}\rho_{\mathbf{n},\mathbf{m}}\,\ket{\mathbf{n}}\bra{\mathbf{m}}, \nonumber \\ 
\rho_{\mathbf{n},\mathbf{m}} &\equiv \rho_{n_{+},n_{-};\,m_{+},m_{-}} \nonumber \\
&= \bra{\mathbf{n}}\hat{\rho}\ket{\mathbf{m}} \equiv \braket{n_{+},n_{-}|\hat{\rho}|m_{+},m_{-}} ,
\label{eq:rho-expansion}
\end{align}
and in the Markovian approximation it evolves according to the Gorini--Kossakowski--Sudarshan--Lindblad (GKSL) master equation \cite{1976Lindblad_CMP,1976Gorini_JMP,2007Breuer}
\begin{align}
\dot{\hat{\rho}}=-\,\mathrm{i}[\hat{H}_\mathrm{s},\hat{\rho}] + \mathcal{L}\hat{\rho} ,
\label{eq:manypol-master}
\end{align}
where the Liouvillian $\mathcal{L}\hat{\rho} = \sum_k \mathcal{D}_k\hat{\rho}$ is defined as the sum of dissipators
\begin{align}
\mathcal{D}_\mathrm{leak}\hat{\rho} &=\sum_{k=\pm}\Gamma_k\!\left(\hat{B}_k\,\hat{\rho}\,\hat{B}_k^\dagger
-\tfrac12\{\hat{B}_k^\dagger\hat{B}_k,\hat{\rho}\}\right),
\label{eq:many-leak}\\
\mathcal{D}_{\downarrow}\hat{\rho}
&=\gamma_\downarrow\!\left(\hat{B}_{-}^\dagger\hat{B}_{+}\,\hat{\rho}\,\hat{B}_{+}^\dagger\hat{B}_{-}
-\tfrac12\{\hat{B}_{+}^\dagger\hat{B}_{-}\hat{B}_{-}^\dagger\hat{B}_{+},\hat{\rho}\}\right),
\label{eq:many-down}\\
\mathcal{D}_{\uparrow}\hat{\rho}
&=\gamma_\uparrow\!\left(\hat{B}_{+}^\dagger\hat{B}_{-}\,\hat{\rho}\,\hat{B}_{-}^\dagger\hat{B}_{+}
-\tfrac12\{\hat{B}_{-}^\dagger\hat{B}_{+}\hat{B}_{+}^\dagger\hat{B}_{-},\hat{\rho}\}\right),
\label{eq:many-up}\\
\mathcal{D}_{\phi}\hat{\rho}
&=\sum_{k=\pm}\gamma_{\phi,k}\!\left(\hat{N}_k\,\hat{\rho}\,\hat{N}_k-\tfrac12\{\hat{N}_k^2,\hat{\rho}\}\right) ,
\label{eq:many-deph}
\end{align}
which describe, respectively, the modes absorption ($\mathcal{D}_\mathrm{leak}$), the UP$\leftrightarrow$LP relaxation/excitation processes ($\mathcal{D}_{\downarrow,\uparrow}$), and dephasing ($\mathcal{D}_{\phi}$). 
Here $\hat{N}_k=\hat{B}_k^\dagger\hat{B}_k$, while the rates $\Gamma_k=-2\,\Im\mathcal S(\omega_k)\ge0$ are fixed by the retarded self-energy $\mathcal S(\omega)=\Sigma(\omega)-\mathrm{i}\Gamma(\omega)/2$ (Sec.~\ref{s:Dyson}); at crossing and in the Drude–Lorentz description of susceptibility $\chi$, one has $\Gamma_{+}=\Gamma_{-} = \gamma_\mathrm{D}/2$, denoted simply as $\Gamma$. Moreover, if the bath is thermal at inverse temperature $\beta$, the detailed balance imposes $\gamma_\uparrow = \gamma_\downarrow \exp\!\big[-\beta(\omega_{+}-\omega_{-})\big]$. It is recalled that the detailed balance condition applies only to the internal UP$\leftrightarrow$LP scattering channel, not to the combined dynamics.

\section{Dissipative dynamics of polariton populations and coherences}
\label{s:dissipative_dynamics}

In Sec.~\ref{s:manypol_pop}, we work in the full number basis $\ket{n_{+}, n_{-}},\,\,n_\pm = 0, 1, 2,\ldots$ (including the vacuum) and derive exact equations of motion for the populations $p_{\mathbf{n}} = \rho_{\mathbf{n},\mathbf{n}}$ and coherences $\rho_{\mathbf{n},\mathbf{m}},\, \mathbf{n} \ne \mathbf{m}$, within the GKSL-Markovian approximations and the assumed structure of jump operators. Under the considered approximations, these equations are exact and preserve both the trace and positivity of the density operator.

In Sec.~\ref{s:drive_rot_frame}, we move to a complementary, coarse-grained (mean-field) description, introducing the coherent amplitudes $A_\pm = \langle B_\pm \rangle$, the branch populations $n_\pm = \langle \hat{N}_\pm \rangle$, and the interbranch coherence $C = \langle \hat{B}_{+}^\dagger \hat{B}_{-} \rangle$. In this description, the GKSL generator reduces to a closed set of nonlinear equations for $(A_\pm, n_\pm, C)$.

To make the developed formalism ready for use in experimental contexts aimed at characterizing the absorption, UP-LP scattering, and dephasing rates, we calculate the time evolution of populations and coherences under coherent continuous wave (CW) drive with complex amplitude $f_\pm$ at the cavity frequency $\omega_\mathrm{c}$, which populates the hybrid modes $\ket{n_\pm}$ (we do not include here the possible Langevin noise \cite{2021Hanson_JOSAB}). Moreover, we also considered the subsequent application of an additional CW low-frequency drive at a frequency $\omega_\mathrm{ext} \approx \omega_{+} - \omega_{-}$ with effective Raman-like coupling strength $\Delta$ (since we set $\hbar=1$, $f_\pm$ and $\Delta$ have the dimensions of a rate). This in principle allows for exploring -- and experimentally measuring, \textit{e.g.}, by pump-probe experiments -- the rates $\Gamma$, $\gamma_{\uparrow,\downarrow}$, and $\gamma_{\phi,\pm}$. 

\subsection{Population and coherence dynamics}
\label{s:manypol_pop}

Using the GKSL master equation, one obtains the equation for the evolution of populations (details of the derivation are given in Appendices~\ref{app:popul_explicit} and \ref{app:popul_explicit_detail}),
\begin{align}
\dot p_{\mathbf{n}}
&= \Gamma\!\Big[
(n_{+}+1)\,p_{n_{+}+1,n_{-}}
+(n_{-}+1)\,p_{n_{+},n_{-}+1}  \nonumber \nonumber  \\
&- (n_{+} + n_{-})\,p_{\mathbf{n}} \Big] \nonumber \\  
&+ (n_{+}+1)n_{-}
\left(\gamma_\downarrow\,p_{n_{+}+1,n_{-}-1}
      -\gamma_\uparrow\,p_{\mathbf{n}}\right)
\nonumber\\
&+ n_{+}(n_{-}\!+\!1)
\left(\gamma_\uparrow\,p_{n_{+}\!-\!1,n_{-}\!+\!1}
      -\gamma_\downarrow\,p_{\mathbf{n}}\right) .
\label{eq:manypol-popEqs}
\end{align}
Equation~\eqref{eq:manypol-popEqs} describes leakage of each polariton at rate $\Gamma$ and UP$\leftrightarrow$LP scattering at rates $\gamma_\downarrow$ and $\gamma_\uparrow$. It is understood that $p_{n_{+},n_{-}} = 0$ whenever an index is negative.

The evolution of the coherences $\rho_{\mathbf{n},\mathbf{m}}$ ($\mathbf{n} \neq \mathbf{m}$) follows from the same GKSL equation as
\begin{widetext}
\begin{align}
\dot\rho_{\mathbf{n},\mathbf{m}}
&= -i(E_{\mathbf{n}}-E_{\mathbf{m}})\,\rho_{\mathbf{n},\mathbf{m}}
\nonumber\\
&\quad +\Gamma\left[\sqrt{(n_{+}+1)(m_{+}+1)}\;
\rho_{\,n_{+}+1,\,n_{-} ;\,m_{+}+1,\,m_{-}}
-\tfrac12(n_{+}+m_{+})\,\rho_{\mathbf{n},\mathbf{m}}\right] \nonumber\\
&\quad +\Gamma\left[\sqrt{(n_{-}+1)(m_{-}+1)}\;
\rho_{\,n_{+},\,n_{-}+1 ;\,m_{+},\,m_{-}+1}
-\tfrac12(n_{-}+m_{-})\,\rho_{\mathbf{n},\mathbf{m}}\right] \nonumber\\
&\quad +\gamma_\downarrow\!\left[
\sqrt{(n_{+}+1)\,n_{-}\, (m_{+}+1)\,m_{-}}\;
\rho_{\,n_{+}+1,\,n_{-}-1 ;\,m_{+}+1,\,m_{-}-1} -\tfrac12\big(n_{+}(n_{-}+1)+m_{+}(m_{-}+1)\big)\rho_{\mathbf{n},\mathbf{m}}\right] \nonumber\\
&\quad +\gamma_\uparrow\!\left[
\sqrt{n_{-}\,(n_{+}+1)\, m_{-}\,(m_{+}+1)}\;
\rho_{\,n_{+}-1,\,n_{-}+1 ;\,m_{+}-1,\,m_{-}+1} -\tfrac12\big(n_{-}(n_{+}+1)+m_{-}(m_{+}+1)\big)\rho_{\mathbf{n},\mathbf{m}}\right] \nonumber\\
&\quad -\tfrac12\sum_{k=\pm}\gamma_{\phi,k}\,(n_k-m_k)^2\,\rho_{\mathbf{n},\mathbf{m}}.
\label{eq:manypol-cohEqs}
\end{align}
\end{widetext}
Importantly, the total trace $\mathrm{Tr}(\hat{\rho})$ is conserved, as required by the Lindblad formalism. A proof is provided in Appendix\,\ref{app:trace}.
Moreover, it can be seen that the dephasing rates $\gamma_{\phi,\pm}$ do not affect the populations and appear only in the time evolution of the coherences. The structure of Eq.~\eqref{eq:manypol-cohEqs} is standard: the first line describes free rotation at the frequency $E_{\mathbf{n}}-E_{\mathbf{m}}$; the second and third lines describe leakage-induced damping; the fourth and fifth lines describe UP$\leftrightarrow$LP transitions; and the last line gives pure dephasing suppression proportional to the number mismatch $(n_k-m_k)^2$. In the single-excitation truncation ($n_{+}+n_{-}\le1$), one recovers the familiar single-polariton optical Bloch equations \cite{1992CohenTannoudji}.

\subsection{Mean-field approximation, coherent drive and interbranch dynamics}
\label{s:drive_rot_frame}
%
Throughout this section, we use a linearized mean-field description in which the Lindblad jump operators are expanded to first order in the polariton occupation numbers. The dynamical equations used here correspond to a linearized, first-moment mean-field closure appropriate for the low-density regime, where occupation
numbers satisfy $\langle N_\pm\rangle \ll 1$ and stimulated scattering terms proportional to $N_{+}N_{-}$ can be neglected. This form of mean-field
approximation is standard in quantum-optical treatments of weakly excited bosonic modes and does not require large populations. Making a concrete example, when calculating mean-field approximations like $\frac{\mathrm{d}}{\mathrm{d}t}{\langle \hat{N}_\pm\rangle}$, terms like $\langle \hat{N}_{+}\hat{N}_{-}\rangle$ arise, which could be safely approximated by $n_{+}n_{n}$. Although they represent stimulated emission and their inclusion could be important, we can discard them compared to terms proportional to $n_\pm$ in the low-density regime. The resulting equations for the time-evolution of populations and coherences are linear in the coherent amplitudes, but can be solved for arbitrary drive strengths. Going beyond the linear-dissipation regime requires keeping the full bosonic excitation factors in the UP$\leftrightarrow$LP Lindblad terms, which introduces nonlinear damping, density-dependent relaxation rates, and modifies the quench rate of UP–LP oscillations. These effects naturally emerge from the jump operators and become relevant at higher excitation densities. 

\cref{eq:manypol-popEqs} allows to evaluate the evolution of the occupation numbers $\langle \hat{N}_\pm \rangle$ according to $\frac{\mathrm{d}}{\mathrm{d}t}{\langle \hat{N}_\pm\rangle} = \mathrm{Tr}(\hat{N}_\pm \dot{\hat{\rho}})$. Indicating $\langle \hat{N}_\pm \rangle$ with $n_\pm$, we have (details of the calculations are given in the Appendix\,\ref{app:average_populations})
\begin{equation}
\frac{\mathrm{d}}{\mathrm{d}t} n_\pm = -\Gamma n_\pm \pm \left(\gamma_{\uparrow} n_{-} - \gamma_{\downarrow} n_{+}\right) .
\label{eq:ntotal-first}
\end{equation}
\cref{eq:ntotal-first}, as derived from the GKSL equation, is exact at the level of first moments for the given choice of jump operators. Only in case we wanted to go beyond the linear model, additional terms $\pm(\gamma_\downarrow-\gamma_\uparrow) n_{+} n_{-}$ would be retained in going from \cref{eq:manypol-popEqs} to \cref{eq:ntotal-first}, making it nonlinear in the occupation number. Such terms can be considered stimulated emission terms: the more quanta are present in one branch, the more efficiently the other branch is fed. They will be considered and presented in a future work.

At all density regimes, however, \cref{eq:ntotal-first} leads to the irreversible decrease of the polaritonic populations as
\begin{equation}
    \frac{\mathrm{d}}{\mathrm{d}t}\,\langle \hat{N}_{+} + \hat{N}_{-}\rangle = \frac{\mathrm{d}}{\mathrm{d}t}\,( n_{+} + n_{-}) =  -\,\Gamma\left(n_{+} +  n_{-}\right) ,
\label{eq:manypol-ntot}
\end{equation}
showing that that UP$\leftrightarrow$LP processes conserve the total polariton number, while leakage removes quanta at a rate $\Gamma$ per polariton. 

A coherent drive with complex amplitude $f_\pm$ at the bare cavity frequency $\omega_\mathrm{c}$ is modeled by
\begin{equation}
H_{\mathrm{drive}}(t)
= \sum_{k=\pm} f_ke^{-\mathrm{i}\omega_\mathrm{c} t}B_k^\dagger + \mathrm{H.c.},
\label{eq:Hdrive_time}
\end{equation}
(H.c standing for Hermitian conjugate), whereas another coherent drive at rate $\Delta$ and frequency $\omega_\mathrm{ext}$ close to $\omega_{+}-\omega_{-}$ couples the UP and LP polaritonic branches according to 
\begin{equation}
\hat{H}_{\mathrm{UP-LP}}(t)
= \Delta \left(\mathrm{e}^{-\mathrm{i}\omega_\mathrm{ext}t}\hat{B}_{+}^\dagger \hat{B}_{-} + \mathrm{e}^{\mathrm{i}\omega_\mathrm{ext}t}\hat{B}_{-}^\dagger \hat{B}_{+}\right). 
\label{eq:H_UPLP}
\end{equation}
Transforming to a frame rotating at $\omega_\mathrm{c}$ by means of a unitary operator
$\hat{U}_\mathrm{c}(t)=e^{\mathrm{i}\omega_\mathrm{c} t(\hat{N}_{+}+\hat{N}_{-})}$, and then writing $\hat{H}_{\mathrm{UP-LP}}(t)$ in the interaction picture, removes the fast optical phase and makes the CW drives time independent. The effective system Hamiltonian becomes
\begin{align}
\hat{H}_{\mathrm{eff}}
&=\frac{\delta}{2}(\hat{N}_{+}-\hat{N}_{-})
+\Delta(\hat{B}_{+}^\dagger \hat{B}_{-} + \hat{B}_{-}^\dagger \hat{B}_{+}) \nonumber \\
&+ \sum_{k=\pm} f_k\hat{B}_k^\dagger 
+ \mathrm{H.c.},
\label{eq:Heff_c_rot_rewritten}
\end{align}
where $\delta=(\omega_{+}-\omega_{-})-\omega_{\mathrm{ext}}$ is the detuning of $\omega_{\mathrm{ext}}$ from the UP-LP frequency difference.

The dissipative dynamics is generated by the GKSL Liouvillian $\mathcal{L}\hat{\rho}$. 
Introducing the average amplitudes $A_\pm=\langle \hat{B}_\pm\rangle$ and using $\frac{\mathrm{d}}{\mathrm{d}t}\langle\hat{B}_\pm \rangle=\mathrm{Tr}(\hat{B}_\pm\dot{\hat{\rho}})$ and the dissipators, the equations for the average amplitudes $A_\pm$ are
\begin{align}
\frac{\mathrm{d}}{\mathrm{d}t} A_{+}
&= -\left(\frac{\Gamma+\gamma_{\phi,+}}{2} + \mathrm{i}\frac{\delta}{2}\right)A_{+}
   - \mathrm{i}\Delta\,A_{-} - \mathrm{i} f_{+},
\label{eq:Aplus_MF_final}\\[4pt]
\frac{\mathrm{d}}{\mathrm{d}t} A_{-}
&= -\left(\frac{\Gamma+\gamma_{\phi,-}}{2} - \mathrm{i}\frac{\delta}{2}\right)A_{-}
   - \mathrm{i}\Delta\,A_{+} - \mathrm{i} f_{-}.
\label{eq:Aminus_MF_final}
\end{align}
By following similar approach, the average of the interbranch coherence $C=\langle \hat{B}_{+}^\dagger \hat{B}_{-}\rangle$ is
\begin{align}
\frac{\mathrm{d}}{\mathrm{d}t} C
&= -\left[\mathrm{i}\delta
+ \Gamma
+ \frac{\gamma_\downarrow+\gamma_\uparrow}{2}
+ \frac{\gamma_{\phi,+}+\gamma_{\phi,-}}{2}\right] C
\nonumber\\
&\quad + \mathrm{i}\Delta\,(n_{-} - n_{+}).
\label{eq:C_MF_final}
\end{align}
The population equations follow from $\dot{n}_\pm =\langle \dot B_\pm^\dagger B_\pm
+ B_k^\dagger\dot B_k\rangle$, which, exploiting \cref{eq:Aplus_MF_final,eq:Aminus_MF_final,eq:C_MF_final}, follow as
\begin{align}
\frac{\mathrm{d}}{\mathrm{d}t} n_\pm  &= -\Gamma n_\pm \pm \left(\gamma_{\uparrow} n_{-} - \gamma_{\downarrow} 
n_{+}\right) \nonumber \\
&\quad \pm 2\Delta\,\Im C - 2\,\Im(f_\pm^* A_\pm)  .
\label{eq:nplusminus_meanfield_rewritten}
\end{align}
\begin{figure*}[!t]
\centerline{
\includegraphics[width=0.33\textwidth]{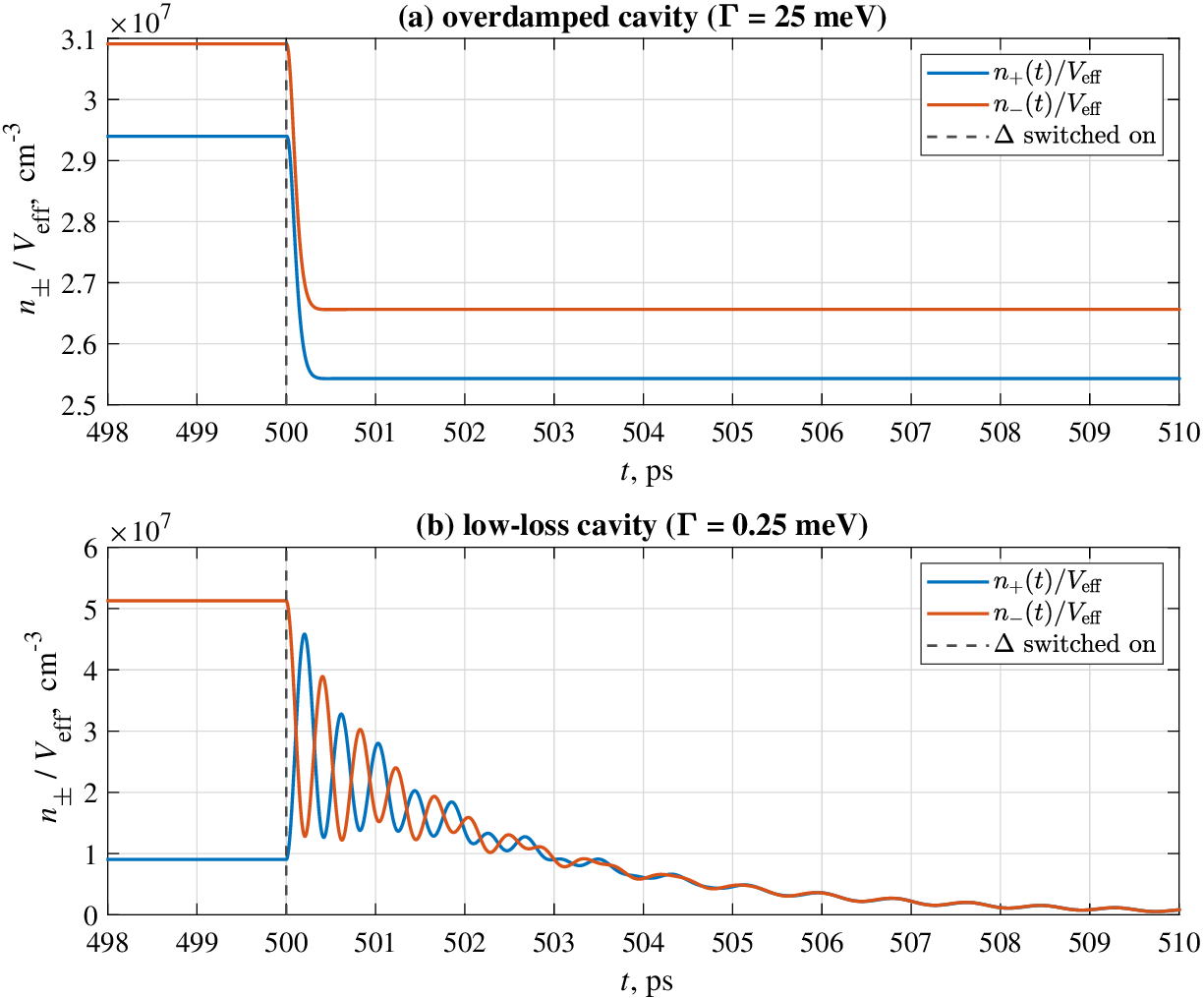}
\includegraphics[width=0.33\textwidth]{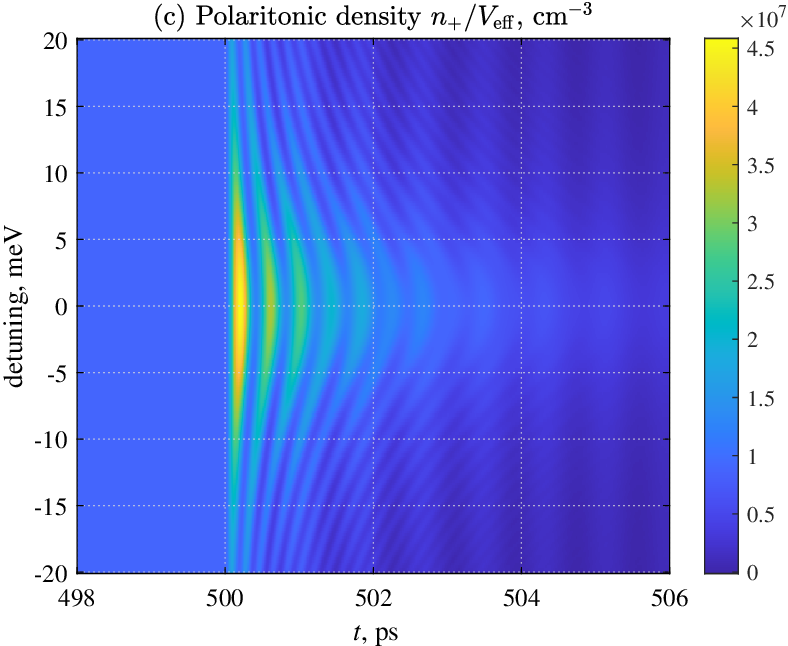}
\includegraphics[width=0.33\textwidth]{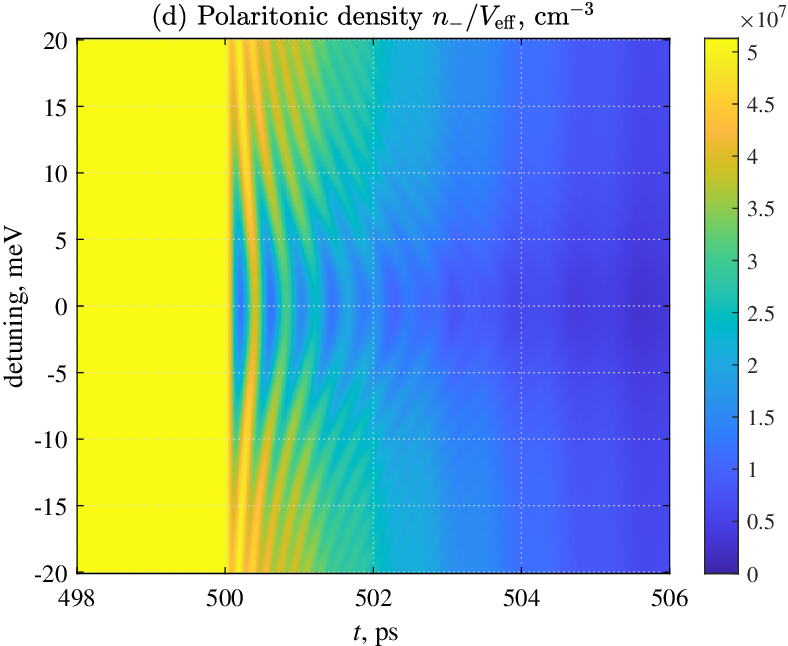}}
\caption{(a) Time evolution of the polaritonic population densities $n_\pm(t) / V_\mathrm{eff}$ for the plasmonic cavity with a gold reflector (overdamped regime, $\Gamma = \gamma_\mathrm{D,\, Au} \approx 25$\,meV, lifetime $\approx 0.026$\,ps). Because $\Gamma \gg \Delta$, turning on the UP--LP coherent coupling at 
$t=500$~ps with zero detuning ($\delta=0$) does not generate oscillatory exchange between the branches: each mode behaves as an independently driven, strongly damped oscillator. 
(b) Same as in (a), but for a low-loss cavity ($\Gamma = 0.25V$\,meV, lifetime $\approx 2.6$\,ps), for which $\Delta \gg \Gamma$ and long-lived relaxation oscillations appear with period $P \approx \pi/\Delta \simeq 0.4$~ps. Panels (c) and (d) show two-dimensional color maps of $n_{+}(t) / V_\mathrm{eff}$ and $n_{-}(t) / V_\mathrm{eff}$ as functions of time and detuning $\delta$, displaying the characteristic Lorentzian profile predicted by \cref{eq:n_ss}.  
For all simulations, 
$\Delta = 5$\,meV ($\approx 7.6$\,ps$^{-1}$) is turned on at $t = 500$\,ps, UP$\rightarrow$LP relaxation rate $\gamma_{\downarrow} = 1$\,ps$^{-1}$, $T = 300$\,K, and $\omega_{+} - \omega_{-} = 0.1$\,eV (yielding $\gamma_{\uparrow} = 0.019$\,ps$^{-1}$). The population density has been evaluated for a $30 \times 30\,\mu$m$^2$ and $1\,\mu$m-thick cavity (effective volume $V_\mathrm{eff} = 9 \times 10^{-10}$\,cm$^{-3}$, just as an example).}
\label{f:fig_4}
\end{figure*}
\cref{eq:nplusminus_meanfield_rewritten} reduces exactly to \cref{eq:ntotal-first} when $\Delta=0$ and $f_\pm=0$.  

To illustrate the dynamics governed by the dissipation rates, we solved the coupled equations \cref{eq:Aplus_MF_final,eq:Aminus_MF_final,eq:C_MF_final,eq:nplusminus_meanfield_rewritten} considering two representative cases: a) the cavity described in Fig.\,\ref{f:fig_1}, which resulted overdamped due to high losses in the metal ($\Gamma = \gamma_\mathrm{D,\,Au}/2 \approx 25$\,meV; b) a low-loss cavity, with $\Gamma = 0.25$\,meV. In both cases, we set a UP$\rightarrow$LP relaxation rate $\gamma_{\downarrow} = 1$\,ps$^{-1}$ and, to avoid additional free parameters, $\gamma_{\phi,\pm} = 0$. The rate $\gamma_{\uparrow}$ was evaluated from $\gamma_{\downarrow}$ for temperature $T = 300$\,K and for $\omega_{+} - \omega_{-} = 0.1$\,eV, which yield $\gamma_{\uparrow} = 0.019$\,ps$^{-1}$.  The coherent drive at $\omega_\mathrm{c}$ with equal amplitudes 
is turned on at $t=0$, and the coherent Raman-like drive with $\Delta = 5$\,meV (or, equivalently $\approx 7.6$\,ps$^{-1}$) is turned on at $t = 500$\,ps. 

Fig.\,\ref{f:fig_4} illustrates the dynamical response of the driven cavity when the coherent Raman-like UP–LP coupling $\Delta$ is switched on at $t = 500$~ps for zero detuning, $\delta=0$. In the overdamped case, Fig.\,\ref{f:fig_4}(a), the condition $\Gamma \gg \Delta$ suppresses the effects of coherent interbranch exchange: each branch behaves as a quasi-independently driven lossy oscillator, and no oscillatory dynamics is observed after the coupling is activated.

In contrast, the low-loss cavity in Fig.\,\ref{f:fig_4}(b) operates in the underdamped regime $\Delta \gg \Gamma, \gamma_{\downarrow}, \gamma_{\uparrow}$. When $\Delta$ is switched on, the modes $\ket{n_\pm}$ hybridize into new normal modes, which display pronounced relaxation oscillations with period $P \approx \pi/\Delta \approx 0.4~\mathrm{ps}$, in agreement with the Rabi-like coupling induced by the low-frequency drive. The oscillation envelope decays over a few picoseconds, consistent with the effective quench rate $\Gamma_{\mathrm{osc}} = \Gamma + \tfrac{3}{4}(\gamma_\downarrow + \gamma_\uparrow)$, which combines irreversible leakage with UP$\rightleftarrows$LP scattering. In fact, in the linear regime and at resonant interbranch driving ($\delta = 0$), the coupled dynamics of the UP–LP population difference $\langle n_{+} - n_{-} \rangle$ and the interbranch coherence follow a two-dimensional Bloch-type equation. Its eigenvalues are
\begin{equation}
    \lambda_\pm = -\left(\Gamma + \frac{3}{4}\left(\gamma_\downarrow+\gamma_\uparrow\right)\right) \pm \mathrm{i} \sqrt{4 \Delta^2 - \frac{\left(\gamma_\downarrow+\gamma_\uparrow\right)^2}{16}},
\end{equation}
so, in the underdamped regime $\Delta > \left(\gamma_\downarrow+\gamma_\uparrow\right)/8$ the UP--LP oscillations decay with an effective rate $\Gamma_{\mathrm{osc}} = \Gamma + \tfrac{3}{4}(\gamma_\downarrow+\gamma_\uparrow)$.

The hybridized normal modes generally have a broader effective linewidth than the original branches. A given drive with interbranch coupling strength $\Delta$ increases the linewidth, reduces the system's susceptibility at the drive frequency, and therefore decreases the net power absorbed from the drive $f_\pm$ (the radiation at $\omega_\mathrm{c}$ that illuminates the cavity). Consequently, the intracavity steady-state population is suppressed by a factor $\Gamma^2/[\Gamma^2+4\Delta^2]$ (if $\delta=0$, for simplicity), which can exceed an order of magnitude in the regime $\Delta\gg\Gamma$.
This result can be obtained by solving the linearized driven-dissipative dynamics in the absence of internal UP$\leftrightarrow$LP relaxation channels, \textit{i.e.}, by setting $\gamma_{\uparrow,\downarrow} = 0$ and neglecting pure dephasing. In this limit, the two polariton branches are driven symmetrically, and the steady-state occupations satisfy $n_{+}=n_{-}$ when $\Delta = 0$ and for $f_{+}=f_{-}$, with a Lorentzian-like spectrum,
\begin{equation}
    n_\pm^\mathrm{ss}(\delta) = |f_\pm|^2\frac{\left(\frac{\Gamma}{2}\right)^2 + \left(\frac{\delta}{2} \pm \Delta\right)^2}{\Big[\left(\frac{\Gamma}{2}\right)^2 + \left(\frac{\delta}{2}\right)^2 + \Delta^2\Big]^2}.
\label{eq:n_ss}
\end{equation}
When the phenomenological UP$\leftrightarrow$LP rates $\gamma_{\uparrow,\downarrow}$ are restored, the steady-state populations become unequal even at $\delta=0$, $\Delta=0$ and equal driving, as Fig.\,\ref{f:fig_4}(a,b) shows, with the ratio $n_{+}/n_{-}$ determined by the rates $\Gamma$ and $\gamma_{\uparrow,\downarrow}$: 
\begin{equation}
    n_\pm^\mathrm{ss,\,general}(\delta) = \frac{\left|-\mathrm{i} f_\pm \left(\kappa_\mp \mp \mathrm{i}\frac{\delta}{2}\right) - \Delta f_\mp \right|^2}{\left|\left(\kappa_{+}+\mathrm{i}\frac{\delta}{2}\right) \left(\kappa_{-}-\mathrm{i}\frac{\delta}{2}\right) + \Delta^2\right|^2} ,
\label{eq:n_ss_general}
\end{equation}
where we defined $\kappa_{+}= \Gamma/2 + \gamma_\downarrow$ and $\kappa_{-}= \Gamma/2 + \gamma_\uparrow$.
The reduction observed when $\Delta$ is turned on is a coherent effect of mode hybridization: the total population still decays only through the irreversible loss rate~$\Gamma$, while $\Delta$ merely redistributes the injected excitations among hybridized modes with a modified decay channel.

In the case of the low-loss cavity, it is $\Delta \gg \Gamma$ and Fig.\,\ref{f:fig_4}(b) shows long-lasting relaxation oscillations with period $P \approx \pi/\Delta  \approx 0.4$\,ps, non present in the panel (a) of the same figure, which refers to the overdamped cavity.

Fig.\,\ref{f:fig_4}(c,d) further show $n_\pm(t)/V_\mathrm{eff}$ as functions of detuning $\delta$, revealing the expected Lorentzian response that peaks at $\delta=0$ and broadens according to the denominator of \cref{eq:n_ss_general} (the population density has been evaluated for a $30 \times 30\,\mu$m$^2$ and $1\,\mu$m-thick cavity, \textit{i.e.}, on an effective volume $V_\mathrm{eff} = 9 \times 10^{-10}$\,cm$^{-3}$, just as an example). The steady-state values $n_\pm^{\mathrm{ss}}(\delta)$ extracted from Fig.\,\ref{f:fig_4} quantitatively match the analytic expression \cref{eq:n_ss_general}, providing a direct means of extracting the cavity loss rate $\Gamma$ from the linewidth of the polaritonic response. Together with an experimental measure of the relaxation oscillations lifetime $\Gamma_\mathrm{osc}^{-1}$, also $\gamma_\downarrow$ can be estimated.

As a final note, it is important to emphasize that in the considered regime the system behaves linearly, so the drive amplitudes $f_\pm$ and $\Delta$ can be large as well as the field-matter coupling strength, provided the induced polariton populations remain small. This condition is naturally satisfied in plasmonic cavities, where leakage usually prevents significant population buildup.

\section{Conclusions}
\label{s:conclusions}

We have presented a unified description of hybrid plasmonic cavities as quantum open systems, combining the Hopfield diagonalization of the lossless Hamiltonian with a propagator renormalization based on the Dyson equation and the Drude–Lorentz susceptibility of the metallic reflector. The resulting complex self-energy $S(\omega) = \Sigma(\omega) - i\Gamma(\omega)/2$ encodes both the frequency renormalization and the irreversible damping of the hybrid plasmon-photon modes, capturing the microscopic origin of material losses due to modes absorption in the metal.

Motivated by tracing out the electromagnetic and electronic environments, we model the dynamics by a GKSL generator that governs the populations and coherences of the upper and lower polariton branches. The Liouvillian includes leakage through $\Gamma = -2\mathrm{Im}\,S(\omega)$, UP$\to$LP and LP$\to$UP internal scattering channels $\gamma_{\downarrow,\uparrow}$, and optional pure dephasing, yielding a closed evolution equation for the average polaritonic occupations. The general solution is exact within the GKSL-Markovian approximations and the assumed structure of jump operators, and fully consistent with the chosen microscopic $\chi(\omega)$ for material losses. In a mean-field picture, we obtained a tractable set of equations for the coherent amplitudes, branch populations, and interbranch coherence, enabling an analytic treatment of steady-state properties.

In the linear regime, but allowing for arbitrary interbranch coupling, field-matter coupling and detuning, the driven steady-state populations follow \cref{eq:n_ss_general}, which generalizes the Lorentzian response formula of \cref{eq:n_ss} by incorporating internal UP$\leftrightarrow$LP scattering and asymmetric decay rates. The quench rate of the UP–LP relaxation oscillations, observed when the Raman-like interbranch coupling is activated, is found analytically to be $\Gamma_{\mathrm{osc}} = \Gamma + \tfrac{3}{4}(\gamma_\downarrow + \gamma_\uparrow)$, a simple expression derived from the eigenvalues of the linearized two-dimensional Bloch-type dynamics. These predictions are in quantitative agreement with the time-domain simulations in Fig. 4, which distinguish overdamped and underdamped regimes depending on the relative magnitudes of $\Gamma$, $\gamma_{\downarrow,\uparrow}$, and the coherent coupling strength.

Going beyond the linear-dissipation regime requires keeping the full bosonic excitation factors in the UP$\leftrightarrow$LP Lindblad terms. Their inclusion will be presented in future work, together with a description of non-Markovian environments.

Finally, we note that the present GKSL formulation can be extended to include explicit Langevin noise forces associated with cavity leakage and material absorption. In the Heisenberg picture, each Lindblad channel generates a corresponding quantum noise operator that drives fluctuations in the polariton amplitudes and enables a stochastic description of the UP and LP fields. Although we focused here on the deterministic evolution of the first moments $\langle\hat{N}_\pm\rangle$, the Langevin formulation provides a natural route to model intensity noise, spectral diffusion, and fluctuation–dissipation relations in plasmonic nanocavities. Incorporating these noise correlations in the polariton basis, and connecting them to the microscopic self-energy of the metallic bath, is a promising direction for future work.

Altogether, the framework developed here provides a minimal yet self-consistent connection between microscopic material response, Green's function electrodynamics, and open quantum system dynamics. It offers a transparent approach to modeling dissipative polariton physics in plasmonic, excitonic, and dielectric cavities, and is directly applicable to pump-probe measurements, THz/microwave spectroscopy of UP–LP transitions, and linewidth engineering in strongly coupled nanophotonic platforms.

\section{Acknowledgments}

This work was supported in part by the European Union under two initiatives of the Italian National Recovery and Resilience Plan (NRRP) of NextGenerationEU: the partnership on Telecommunications of the Future (PE00000001 – program “RESTART”), and the National Centre for HPC, Big Data and Quantum Computing (CN00000013 – CUP E13C22000990001).


\appendix

\section{Operator form of the matrix elements in the polariton basis}
\label{app:popul_explicit}

In this appendix, we present the explicit evaluation of some relations that allow for the evaluation of any GKSL term in closed and explicit forms, followed by the explicit derivation of the time evolution of populations (the diagonal coefficients of the density matrix).

Starting from the definitions of Fock states $\ket{\mathbf{n}} \equiv \ket{n_{+},n_{-}}$ generated from the vacuum according to \cref{eq:Fock_states} in the main text, the dyadic operator
\begin{equation}
    \hat{P}_{\mathbf{n},\mathbf{m}} \equiv \ket{\mathbf{n}}\bra{\mathbf{m}}
\end{equation}
provides a convenient operator basis. The density operator then reads
\begin{equation}
\hat\rho=\sum_{\mathbf{n},\mathbf{m}}\rho_{\mathbf{n},\mathbf{m}}\,\hat{P}_{\mathbf{n},\mathbf{m}},
\qquad \rho_{\mathbf{n},\mathbf{m}}=\bra{\mathbf{n}}\hat\rho\ket{\mathbf{m}}
\label{eqapp:rho-expansion}
\end{equation}
which can be written equivalently according to \cref{eq:rho-expansion}. From these definitions and from the operator number $\hat{{N}_k} = \hat{B}_k^\dagger \hat{B}_k$, the following useful identities follow:
\begin{align}
\hat{N}_k\,\hat{P}_{\mathbf{n},\mathbf{m}} &= n_k\,\hat{P}_{\mathbf{n},\mathbf{m}},\qquad
\hat{P}_{\mathbf{n},\mathbf{m}}\,\hat{N}_k = m_k\,\hat{P}_{\mathbf{n},\mathbf{m}}, \\
\hat{B}_k\,\hat{P}_{\mathbf{n},\mathbf{m}} &= \sqrt{n_k}\,\hat{P}_{\mathbf{n}-\mathbf{e}_k,\ \mathbf{m}} \\
\hat{B}_k^\dagger\,\hat{P}_{\mathbf{n},\mathbf{m}} &= \sqrt{n_k+1}\,\hat{P}_{\mathbf{n}+\mathbf{e}_k,\ \mathbf{m}},\\
\hat{P}_{\mathbf{n},\mathbf{m}}\,\hat{B}_k &= \sqrt{m_k}\,\hat{P}_{\mathbf{n} ,\ \mathbf{m}-\mathbf{e}_k}, \\
\hat{P}_{\mathbf{n},\mathbf{m}} \,\hat{B}_k^\dagger &= \sqrt{m_k+1}\,\hat{P}_{\mathbf{n} ,\ \mathbf{m}+\mathbf{e}_k}.
\end{align}
Consequently, we have, for example
\begin{align}
\hat{B}_k\,\hat{P}_{\mathbf{n},\mathbf{m}}\,\hat{B}_k^\dagger
&= \sqrt{n_k m_k}\;\hat{P}_{\mathbf{n}-\mathbf{e}_k,\ \mathbf{m}-\mathbf{e}_k},\\
\hat{B}_{-}^\dagger\hat{B}_{+}\,\hat{P}_{\mathbf{n},\mathbf{m}}\,\hat{B}_{+}^\dagger\hat{B}_{-}
&= \sqrt{(n_{+}+1)n_{-}\, (m_{+}+1)m_{-}} \nonumber \\
&\quad \times \hat{P}_{\mathbf{n}+\mathbf{e}_{+}-\mathbf{e}_{-},\ \mathbf{m}+\mathbf{e}_{+}-\mathbf{e}_{-}},\\
\hat{B}_{+}^\dagger\hat{B}_{-}\,\hat{P}_{\mathbf{n},\mathbf{m}}\,\hat{B}_{-}^\dagger\hat{B}_{+}
&= \sqrt{n_{-}(n_{+}+1)\, m_{-}(m_{+}+1)} \nonumber \\
&\quad \times \hat{P}_{\mathbf{n}-\mathbf{e}_{+}+\mathbf{e}_{-},\ \mathbf{m}-\mathbf{e}_{+}+\mathbf{e}_{-}}.
\end{align}
These relations provide any GKSL term in closed form. For example,
\begin{align}
    \hat{B}_k\,\hat\rho\,\hat{B}_k^\dagger &= \sum_{\mathbf{n},\mathbf{m}}\rho_{\mathbf{n},\mathbf{m}}\,
\sqrt{n_k m_k}\;\hat{P}_{\mathbf{n}-\mathbf{e}_k,\ \mathbf{m}-\mathbf{e}_k} \\
\{\hat{B}_k^\dagger\hat{B}_k,\hat\rho\} &= \sum_{\mathbf{n},\mathbf{m}}(n_k+m_k)\,\rho_{\mathbf{n},\mathbf{m}}\,\hat{P}_{\mathbf{n},\mathbf{m}}.
\end{align}

Together with \eqref{eqapp:rho-expansion} this reproduces the element-wise evolution equations of populations and coherences.

\section{Populations: explicit calculations}
\label{app:popul_explicit_detail}

Let us consider the elements of \cref{eq:manypol-master} that belong to populations, i.e., $\dot{\hat{\rho}}_{\mathbf{n}, \mathbf{n}}$. 
The Hamiltonian operator is diagonal in the polaritonic basis, $\hat{H}_\mathrm{s}=\sum_\mathbf{p} E_\mathbf{p}\,\hat{P}_{\mathbf{p} \mathbf{p}}$, hence
\begin{align}
\hat{H}_\mathrm{s}\,\hat{P}_{\mathbf{n}\mathbf{m}}
&=\sum_{\mathbf{p}} E_{\mathbf{p}}\,\hat{P}_{\mathbf{p}\mathbf{p}}\hat{P}_{\mathbf{n}\mathbf{m}}
=\sum_{\mathbf{p}} E_{\mathbf{p}}\,\delta_{\mathbf{p},\mathbf{n}}\,\hat{P}_{\mathbf{p}\mathbf{m}}
=E_{\mathbf{n}}\,\hat{P}_{\mathbf{n}\mathbf{m}},\\
\hat{P}_{\mathbf{n}\mathbf{m}}\,\hat{H}_\mathrm{s}
&=\sum_{\mathbf{p}} E_{\mathbf{p}}\,\hat{P}_{\mathbf{n}\mathbf{m}}\hat{P}_{\mathbf{p}\mathbf{p}}
=\sum_{\mathbf{p}} E_{\mathbf{p}}\,\delta_{\mathbf{p},\mathbf{m}}\,\hat{P}_{\mathbf{n}\mathbf{p}}
=E_{\mathbf{m}}\,\hat{P}_{\mathbf{n}\mathbf{m}},
\end{align}
from which the general identity $\big[\hat{H}_\mathrm{s},\hat{P}_{\mathbf{n}\mathbf{m}}\big]
=(E_{\mathbf{n}}-E_{\mathbf{m}})\,\hat{P}_{\mathbf{n}\mathbf{m}}$ follows.
For $\mathbf{n}=\mathbf{m}$, it is $\big[\hat{H}_\mathrm{s},\hat{P}_{\mathbf{n}\mathbf{n}}\big] =(E_{\mathbf{n}}-E_{\mathbf{n}})\,\hat{P}_{\mathbf{n}\mathbf{n}}=0$, therefore the commutator $\big[\hat{H}_\mathrm{s},\hat\rho\big]$ leaves populations $\rho_{\mathbf{n},\mathbf{n}}$ unchanged.

More generally, for a density operator $\hat\rho=\sum_{\mathbf{n},\mathbf{m}}\rho_{\mathbf{n},\mathbf{m}}\hat{P}_{\mathbf{n}\mathbf{m}}$, it is $\big[\hat{H}_\mathrm{s},\hat\rho\big] =\sum_{\mathbf{n},\mathbf{m}}\rho_{\mathbf{n},\mathbf{m}}\,(E_{\mathbf{n}}-E_{\mathbf{m}})\,\hat{P}_{\mathbf{n}\mathbf{m}}$, imparting phase rotations only to coherences $\rho_{\mathbf{n},\mathbf{m}}$, with $\mathbf{n}\neq\mathbf{m}$.  If $E_{\mathbf{n}}=E_{\mathbf{m}}$ are degenerate, that coherence is also stationary under $\hat{H}_\mathrm{s}$.


Concerning the dissipators, it is convenient to define the diagonal coefficients of the density matrix $\rho_{\mathbf{n},\mathbf{n}} = \bra{\mathbf{n}}\hat{\rho}\ket{\mathbf{n}} \equiv \braket{n_{+},n_{-}|\hat{\rho}|n_{+},n_{-}}$ shortly as $p_{\mathbf{n}} \equiv p_{n_{+},n_{-}}$. We have, for example:
\begin{align}
&\mathcal{D}_\mathrm{leak}\hat{\rho}_{\mathbf{n},\mathbf{n}} = \Gamma \, \sum_{k=\pm}\left(\hat{B}_k\,\hat{\rho}\,\hat{B}_k^\dagger
-\tfrac12\{\hat{B}_k^\dagger\hat{B}_k,\hat{\rho}\}\right)   \nonumber \\
&\quad= \Gamma \, \sum_{k=\pm}  \left(p_{\mathbf{n}} \,
n_k\;\hat{P}_{\mathbf{n}-\mathbf{e}_k,\ \mathbf{n}-\mathbf{e}_k} \right. \nonumber \\ 
&\left.\qquad - \frac{1}{2} p_{\mathbf{n}} \, (n_k+n_k)\,\hat{P}_{\mathbf{n},\mathbf{n}} \right)\nonumber \\
&\quad= \Gamma \, \sum_{k=\pm}  \left(p_{\mathbf{n}+\mathbf{e}_k} \,
(n_k+1)\; -  p_{\mathbf{n}} \, n_k\,\, \right) \, \hat{P}_{\mathbf{n},\mathbf{n}}\nonumber \\
&\quad= \Gamma \, \left(p_{n_{+}+1,\,n_{-}} \,
(n_{+}+1) -  p_{n_{+},\,n_{-}} \, n_{+}\right)\, \hat{P}_{\mathbf{n},\mathbf{n}} \nonumber \\
&\quad + \Gamma \, \left(p_{n_{+},\,n_{-}+1} 
(n_{-}+1) -  p_{n_{+},\,n_{-}} \, n_{-}\,\, \right)\, \hat{P}_{\mathbf{n},\mathbf{n}} .
\end{align}
Therefore, for each $\mathbf{n}$, this dissipator contributes with a coefficient
\begin{align}
[\mathcal{D}_\mathrm{leak}\hat{\rho}]_\mathbf{n} &= \Gamma \, \Big[p_{n_{+}+1,\,n_{-}} \,
(n_{+}+1) + p_{n_{+},\,n_{-}+1} (n_{-}+1) \nonumber \\
&-  p_{n_{+},\,n_{-}} \, (n_{+}+n_{-})\Big] .
\end{align}
The explicit expressions of the other dissipators can be obtained following a similar path.

\section{Trace preservation: operator and component proofs}
\label{app:trace}

Let $\mathrm{Tr}\,\hat\rho(t)=\sum_{n_{+},n_{-}} p_{n_{+},n_{-}}(t)$, where
$p_{n_{+},n_{-}}\equiv \rho_{(n_{+},n_{-}),(n_{+},n_{-})}$ are the diagonal elements in the
polariton Fock basis. Using the cyclicity of the trace and $\mathrm{Tr}(\, [H_\mathrm{s},\rho]\,)=0$, we find
\begin{align}
&\frac{\mathrm{d}}{\mathrm{d}t}\mathrm{Tr}\,\rho
=\sum_j \mathrm{Tr}\!\left(\mathcal{D}[L_j]\rho\right) \nonumber\\
&\quad=\sum_j \Big(\mathrm{Tr}[L_j\rho L_j^\dagger]
-\tfrac12\mathrm{Tr}[L_j^\dagger L_j \rho]
-\tfrac12\mathrm{Tr}[\rho L_j^\dagger L_j]\Big)\nonumber\\
&\quad=\sum_j \Big(\mathrm{Tr}[L_j^\dagger L_j \rho]
-\tfrac12\mathrm{Tr}[L_j^\dagger L_j \rho]
-\tfrac12\mathrm{Tr}[L_j^\dagger L_j \rho]\Big)=0.
\end{align}
This holds separately for each dissipator: leakage \eqref{eq:many-leak},
relaxation/excitation \eqref{eq:many-down}–\eqref{eq:many-up}, and dephasing
\eqref{eq:many-deph}.

\section{Coherences: explicit calculations}
\label{app:coher_explicit}

This appendix shown the explicit derivation of \cref{eq:manypol-cohEqs}. 
\begin{itemize}
    \item In the Appendix\,\ref{app:popul_explicit}, we have shown that for $\mathbf{n} \ne \mathbf{m}$ it is $\big[\hat{H}_\mathrm{s},\hat\rho\big] =\sum_{\mathbf{n},\mathbf{m}}\rho_{\mathbf{n},\mathbf{m}}\,(E_{\mathbf{n}}-E_{\mathbf{m}})\,\hat{P}_{\mathbf{n}\mathbf{m}} \ne 0$, therefore is is, explicitly,
\begin{align}
    -\mathrm{i}\big[\hat{H}_\mathrm{s},\hat\rho\big] &= -\mathrm{i}\,\big[(n_{+}-m_{+})\omega_{+} + (n_{-} m_{-})\omega_{-}\big]   \nonumber \\
    &\quad \times \rho_{\,n_{+},n_{-};\,m_{+},m_{-}}.
\end{align}
\item{The leakage (modes absorption in the metal) is described by $L_\pm=\sqrt{\Gamma}\,B_\pm$.} Using $\braket{n_{+},n_{-}|\hat{B}_{+}\,|n_{+}{+}1,n_{-}} =\sqrt{n_{+}{+}1}$
and $\braket{m_{+}{+}1,m_{-}|\hat{B}_{+}^\dagger\,|m_{+},m_{-}} = \sqrt{m_{+}{+}1}$ (and similarly for $\hat{B}_{-}$), the related dissipator $\mathcal{D}_\mathrm{leak} \hat{\rho}$ contributes to to coherences in \cref{eq:manypol-master} with the term

\begin{align}
&\Gamma\,\Big[\sqrt{(n_{+}+1)(m_{+}+1)}\;
\rho_{\,n_{+}+1,\,n_{-} ;\,m_{+}+1,\,m_{-}} \Big. \nonumber\\
&\Big.-\tfrac12(n_{+}+m_{+})\,\rho_{\,n_{+},n_{-};\,m_{+},m_{-}}\Big.\nonumber\\
&+\sqrt{(n_{-}+1)(m_{-}+1)}\;
\rho_{\,n_{+},\,n_{-}+1 ;\,m_{+},\,m_{-}+1} \Big.   \nonumber\\
&\Big.-\tfrac12(n_{-}+m_{-})\,\rho_{\,n_{+},n_{-};\,m_{+},m_{-}}\Big].
\end{align}
\item The relaxation UP$\to$LP is determined by $J_\downarrow=\sqrt{\gamma_\downarrow}\,\hat{B}_{-}^\dagger \hat{B}_{+}$, which acts on the Fock states according to $\hat{B}_{-}^\dagger \hat{B}_{+}\ket{n_{+}+1,n_{-}-1}=\sqrt{(n_{+}+1)n_{-}}\,\ket{n_{+},n_{-}}$ and $\bra{m_{+},m_{-}}\hat{B}_{+}^\dagger \hat{B}_{-}=\sqrt{(m_{+}+1)\,m_{-}}\,\bra{m_{+}+1,\,m_{-}-1}$.
Therefore, the related dissipator $\mathcal{D}_\downarrow \hat{\rho}$ contributes to coherences in \cref{eq:manypol-master} with the term 
\begin{align}
&\quad \gamma_\downarrow\Big[
\sqrt{(n_{+}+1)\,n_{-}\, (m_{+}+1)\,m_{-}} \nonumber\\
&\quad\times \rho_{\,n_{+}+1,\,n_{-}-1 ;\,m_{+}+1,\,m_{-}-1} 
\nonumber\\
&-\tfrac12\big(n_{+}(n_{-}+1)+m_{+}(m_{-}+1)\big)\,
\rho_{\,n_{+},n_{-};\,m_{+},m_{-}}\Big].
\end{align}
The excitation LP$\to$UP is determined by $J_\uparrow=\sqrt{\gamma_\uparrow}\,\hat{B}_{+}^\dagger \hat{B}_{-}$, and a similar procedure leads to the related contribution.  
\item The dephasing is determined by $L_{\phi,k}=\sqrt{\gamma_{\phi,k}}\,\hat{N}_k$. Since $\hat{N}_k\ket{n_{+},n_{-}}=n_k\ket{n_{+},n_{-}}$, the contribution coming from the related dissipator is
\begin{equation}
    -\,\tfrac12\sum_{k=\pm}\gamma_{\phi,k}\,(n_k-m_k)^2\;
\rho_{\,n_{+},n_{-};\,m_{+},m_{-}}.
\end{equation}

\end{itemize}


\section{Derivation of the average population equations}
\label{app:average_populations}

We start from the exact population equation obtained from the GKSL dissipators,
\begin{align}
\dot p_{n_+,n_-}
&=
\Gamma\Big[(n_{+}+1)\,p_{n_{+}+1,n_{-}}
+(n_{-}+1)\,p_{n_{+},n_{-}+1} \nonumber \\
&\quad-(n_{+}+n_{-})p_{n_{+},n_{-}}\Big]
\label{eq:app-dotp}
\\
&\quad
+(n_{+}+1)n_{-}
\Big(\gamma_{\downarrow}\,p_{n_{+}+1,n_{-}-1}
     -\gamma_{\uparrow}\,p_{n_{+},n_{-}}\Big)
\nonumber\\
&\quad
+n_{+}(n_{-}+1)
\Big(\gamma_{\uparrow}\,p_{n_{+}-1,n_{-}+1}
     -\gamma_{\downarrow}\,p_{n_{+},n_{-}}\Big),
\nonumber
\end{align}
with the convention that $p_{n_+,n_-}=0$ if any index is negative.  The
expectation values are
\begin{equation}
\langle \hat{N}_{+}\rangle=\sum_{n_+,n_-}n_+\,p_{n_+,n_-},
\qquad
\langle \hat{N}_{-}\rangle=\sum_{n_+,n_-}n_-\,p_{n_+,n_-}.
\label{eq:app-exp}
\end{equation}

We compute $\frac{\mathrm{d}}{\mathrm{d}t}\langle\hat{N}_\pm\rangle=\sum_{\mathbf n} n_\pm\dot p_{\mathbf n}$
by grouping the contributions from (i) leakage ($\Gamma$),
(ii) UP$\!\to$LP relaxation ($\gamma_\downarrow$), and
(iii) LP$\!\to$UP excitation ($\gamma_\uparrow$).  All manipulations follow
from index–shift identities such as
$p_{n_{+}+1,n_-}\mapsto p_{m_+,n_-}$ with $m_+=n_++1$.

\subsection{Leakage}

Using only the $\Gamma$-proportional part of Eq.~\eqref{eq:app-dotp},
\begin{align}
\dot p_{n_+,n_-}\big|_\Gamma
&=\Gamma\Big[(n_{+}+1)p_{n_{+}+1,n_{-}}
+(n_{-}+1)p_{n_{+},n_{-}+1} \nonumber \\
&-(n_{+}+n_{-})p_{n_{+},n_{-}}\Big],    
\end{align}
we obtain
\begin{align}
\frac{\mathrm{d}}{\mathrm{d}t}\langle\hat{N}_\pm\rangle\big|_\Gamma
&=\sum_{n_+,n_-}n_{+}\,\dot p_{n_+,n_-}\big|_\Gamma
\equiv T_1+T_2+T_3 , 
\end{align}
where, shifting the summation indices in the inflow terms, 
\begin{align}
T_1 &= \Gamma\sum_{n_+,n_-}n_{+}(n_{+}+1)p_{n_{+}+1,n_-}
 =\Gamma(\langle \hat{N}_+^2\rangle - \langle \hat{N}_+\rangle),\\
T_2 &= \Gamma\sum_{n_+,n_-}n_{+}(n_-+1)p_{n_+,n_-+1}
=\Gamma\langle \hat{N}_{+}\hat{N}_{-}\rangle,\\
T_3 &= -\Gamma\sum_{n_+,n_-}n_{+}(n_++n_-)p_{n_+,n_-}
=-\Gamma(\langle \hat{N}_+^2\rangle + \langle \hat{N}_{+}\hat{N}_{-} \rangle.
\end{align}
Summing them (and corresponding terms concerning $\langle\hat{N}_{-}\rangle$) gives
\begin{equation}
\frac{\mathrm{d}}{\mathrm{d}t}\langle\hat{N}_\pm\rangle\big|_\Gamma
=-\,\Gamma\,n_\pm, 
\label{eq:app-leak}
\end{equation}

\subsection{UP{\texorpdfstring{$\to$}{→}}LP relaxation and LP{\texorpdfstring{$\to$}{→}}UP excitation}

The remaining terms in Eq.~\eqref{eq:app-dotp},
\begin{equation}
\begin{aligned}
\dot p_{n_+,n_-}\big|_{\downarrow,\uparrow}
&=
(n_{+}+1)n_{-}\big(\gamma_{\downarrow}p_{n_{+}+1,n_{-}-1}
                  -\gamma_{\uparrow}p_{n_{+},n_{-}}\big)
\\
&\quad
+n_{+}(n_{-}+1)\big(\gamma_{\uparrow}p_{n_{+}-1,n_{-}+1}
                   -\gamma_{\downarrow}p_{n_{+},n_{-}}\big),
\end{aligned}
\label{eq:app-relax}
\end{equation}
describe population transfer between the branches.
Each of the four terms is evaluated by index shifts exactly as in the leakage
case.  Although higher–order correlators such as
$\langle \hat{N}^2_{+}\hat{N}_{-}\rangle$ appear at intermediate steps, these cancel out when all inflow and outflow contributions are summed. The result is the closed linear exchange law
\begin{align}
\frac{\mathrm{d}}{\mathrm{d}t}\langle\hat{N}_{+}\rangle\big|_{\downarrow,\uparrow}
&= -(\gamma_\downarrow+\gamma_\uparrow)n_{+}
   +(\gamma_\downarrow+\gamma_\uparrow)n_{-},
\label{eq:app-relax-plus}\\[2pt]
\frac{\mathrm{d}}{\mathrm{d}t}\langle\hat{N}_{-}\rangle\big|_{\downarrow,\uparrow}
&= -(\gamma_\downarrow+\gamma_\uparrow)n_{-}
   +(\gamma_\downarrow+\gamma_\uparrow)n_{+}.
\label{eq:app-relax-minus}
\end{align}

\subsection{Final result}

Summing the leakage \cref{eq:app-leak} and relaxation/excitation \cref{eq:app-relax-plus,eq:app-relax-minus} contributions gives the coupled equations
\begin{align}
\frac{\mathrm{d}}{\mathrm{d}t}\langle\hat{N}_{+}\rangle
&= -(\Gamma+\gamma_{\downarrow})n_{+}
   +\gamma_{\uparrow}n_{-},
\label{eq:app-nplus-final}
\\[4pt]
\frac{\mathrm{d}}{\mathrm{d}t}\langle\hat{N}_{-}\rangle
&= -(\Gamma+\gamma_{\uparrow})n_{-}
   +\gamma_{\downarrow}n_{+}.
\label{eq:app-nminus-final}
\end{align}
Adding these equations shows that UP$\!\leftrightarrow$LP transitions conserve
the total polariton number, while leakage removes excitations at rate~$\Gamma$:
\begin{equation}
\frac{\mathrm{d}}{\mathrm{d}t}\langle \hat{N}_{+}+\hat{N}_{-}\rangle
=-\,\Gamma\,\left( n_{+}+n_{-}\right).
\label{eq:app-ntotal}
\end{equation}
Equations~\eqref{eq:app-nplus-final}–\eqref{eq:app-nminus-final} coincide with \cref{eq:nplusminus_meanfield_rewritten} in the main text when the coherent drive and the interbranch coupling are switched off, confirming the consistency between the many–polariton and mean-field descriptions. 

In the main text, when the meaning is clear we indicated $\langle\hat{N}_\pm\rangle$ more simply with $n_\pm$.


%

\end{document}